\documentclass[%
 aip,
rsi,
 amsmath,amssymb,
 reprint,%
]{revtex4-1}

\usepackage{graphicx}
\usepackage{dcolumn}
\usepackage{bm}

\usepackage[utf8]{inputenc}
\usepackage[T1]{fontenc}
\usepackage{mathptmx}
\usepackage{amsmath}
\usepackage{etoolbox}
\usepackage{csquotes}
\usepackage{amsfonts}
\usepackage{xcolor}
\usepackage{verbatim}
\usepackage{upgreek}

\draft 

\DeclareMathOperator*{\argmin}{arg\,min}
\newcommand{\iu}{\mathrm{i}}
\newcommand{\eu}{\mathrm{e}}

\begin{document}


\title{Polarization-Resolved Transmission Matrices of Specialty Optical Fibers} 



\author{Erin S. Lamb}
 \email{elamb@ofsoptics.com}
 \affiliation{OFS Laboratories, 19 Schoolhouse Road, Somerset, NJ 08873, USA}
 
\author{Tristan Kremp}
\affiliation{OFS Laboratories, 19 Schoolhouse Road, Somerset, NJ 08873, USA}

\author{David J. DiGiovanni}
\affiliation{OFS Laboratories, 19 Schoolhouse Road, Somerset, NJ 08873, USA}

\author{Paul S. Westbrook}
\affiliation{OFS Laboratories, 19 Schoolhouse Road, Somerset, NJ 08873, USA}


\date{\today}

\begin{abstract}
Transmission matrix measurements of multimode fibers are now routinely performed in numerous labs, enabling control of the electric field at the distal end of the fiber and paving the way for the potential application to ultrathin medical endoscopes with high resolution. The same concepts are applicable to other areas such as space division multiplexing, targeted power delivery, fiber laser performance, and the general study of the mode coupling properties of the fiber. However, the process of building an experimental setup and developing the supporting code to measure the fiber's transmission matrix remains challenging and time consuming with full details on experimental design, data collection, and supporting algorithms spread over multiple papers or lacking in detail. Here, we outline a complete and self-contained description of the experiment we use to measure fully polarization-resolved transmission matrices, which enable full control of the electric field, in contrast to the more common scalar setups. Our specific implementation of the full polarization experiment is new and is easy to align while providing flexibility to switch between full-polarization and scalar measurements if desired. We utilize a spatial light modulator to measure the transmission matrix using linear phase gratings to generate the basis functions and measure the distal electric field using phase-shifting interferometry with an independent reference beam derived from the same laser. We introduce a new method to measure and account for the phase and amplitude drift during the measurement using a Levenberg-Marquardt nonlinear fitting algorithm. Finally, we describe creating distal images through the multimode fiber using phase-to-amplitude shaping techniques to construct the correct input electric field through a superposition of the basis functions with the phase-only spatial light modulator. We show that results are insensitive to the choice of phase-to-amplitude shaping technique as quantified by measuring the contrast of a razor blade at the distal end of the fiber, indicating that the simplest but most power efficient method may be the best choice for many applications. We also discuss some of the possible variations on the setup and techniques presented here and highlight the details that we have found key in achieving high fidelity distal control. Throughout the paper, we discuss applications of our setup and measurement process to a variety of specialty fibers, including fibers with harsh environment coatings, coreless fibers, rectangular core fibers, pedestal fibers, and a pump-signal combiner based on a tapered fiber bundle. This demonstrates the usefulness of these techniques across a variety of application areas and shows the flexibility of our setup in studying various fiber types.
\end{abstract}


\maketitle 

\section{Introduction}
\label{sec:introduction}

Transmission matrices (TMs) have emerged as powerful tools for distal control through multimode fibers (MMFs) \cite{Cao2023} and turbid media.\cite{Mosk2012} Through proper phase control at the proximal end of the fiber, typically achieved through the use of a spatial light modulator (SLM) \cite{Zhang2014} or digital micromirror device (DMD),\cite{Song2018} the mode coupling within the MMF can be probed by launching a series of transverse basis functions. With enough of these measurements, the transmission matrix describing propagation through the fiber can be constructed and used to predict the required fiber launch to achieve a desired distal output, such as a spot that can be scanned to perform imaging through the fiber. This technique holds the promise of enabling ultrathin medical endoscopes with smaller form-factor and greater flexibility than coherent fiber bundles and with high resolution and spot scanning without mechanical motion. In addition, the same concepts are more broadly applicable in areas such as space division multiplexing, targeted power delivery, tailoring fiber laser performance, and for studying light propagation in MMF. 

A variety of tools and techniques have been designed to enable TM measurements and control of the electric field through MMFs (and other scattering media). Popoff et al. present an early use of an SLM and interferometric electric field measurement to probe the statistical properties of a thick scattering medium and demonstrate focusing through it. \cite{Popoff2010} Bianchi and Di Leondardo follow with the calibration of a 1-meter long MMF using an SLM to demonstrate holographic manipulation of particles and scanning fluorescence microscopy using a scanned spot at the distal end of their fiber. \cite{Bianchi2012,DiLeonardo2011} \v{C}i\v{z}m\'{a}r and coworkers extend the discussion of calibration of MMFs using SLMs and show the creation of images at the distal end of the MMF; \cite{Cizmar2011} this work was later expanded to utilize GPU computations to increase the speed of the TM measurement and application. \cite{Ploschner2014} Other groups describe using a different technique, digital phase conjugation,\cite{Yariv1978,Vellekoop2007,Mididoddi2020} to also focus and scan a small number of spots through MMFs while requiring launch access from both fiber ends. \cite{Papadopoulos2012,Papadopoulos2013} The digital phase conjugation techniques were expanded by Farahi and coworkers through the use of a distal beacon to allow for some bending compensation using a look-up table approach. \cite{Farahi2013} Carpenter et al. have demonstrated full spatial-temporal characterization of an MMF using an SLM by measuring the TM over a range of wavelengths and showing that they can create any desired spatial output from a fiber for a given polarization and wavelength. \cite{Carpenter2016} Although the basic TM measurement techniques remain the same, DMDs are becoming increasingly popular for these measurements due to their orders-of-magnitude speed increase. \cite{Caravaca2013,Ren2015,Turtaev2017,Hoffmann2018,Zhao2021,Popoff2023,Gutierrez2023} Additionally, it is possible to perform focusing through an MMF using an ultrafast grating light valve \cite{Tzang2019} or through manipulation of the fiber itself. \cite{Resisi2020,Shekel2023} All of the devices used to control the electric field on the proximal side of the fiber must be capable of shaping the light appropriately both for the TM measurement and for creating the tailored electric fields needed to produce the desired output at the distal end of the fiber. Methods of doing this have trade-offs among speed, accuracy, and overall power efficiency. 

Once calibrated, MMFs are being utilized in an increasing variety of applications. Loterie and coworkers demonstrate digital confocal microscopy through an MMF endoscope. \cite{Loterie2015,loterie2015_2} Caravaca-Aguirre and Piestun show that careful selection of the MMF enables fluorescence microscopy through the fiber with some tolerance to fiber movement. \cite{Caravaca2017} Numerous groups have demonstrated ultrathin, TM-calibrated MMF endoscopes to conduct deep brain fluorescence imaging for in-vivo and ex-vivo applications. \cite{Ohayon2018,Turtaev2018,Vasquez2018,Stibuurek2023} Gusachenko and coworkers perform Raman spectroscopic imaging through an MMF. \cite{Gusachenko2017} In the smallest demonstrated probe, \cite{Turtaev2018} the excess cladding glass is removed from the MMF to achieve a probe diameter of 60\,µm for in-vivo studies of the visual cortex and hippocampus of anaesthetised mice. In this and other demonstrations,\cite{Ohayon2018} the calibrated fiber is kept short and used in a needle-like application so that manipulation of the fiber does not degrade the calibration during its use. 

In addition to these applications, multiple groups have applied these MMF endoscopes to nonlinear imaging techniques. Morales-Delgado and coworkers use selective phase conjugation to control the excited modes and deliver 500-fs pulses to the distal end of an MMF, \cite{Morales2015delivery} followed by an example of two-photon imaging using an MMF endoscope. \cite{Morales2015} Sivankutty et al. also demonstrate two-photon imaging through a few centimeters of calibrated rigid graded-index fiber. \cite{Sivankutty2016} Kakkava et al. select fibers where the TM calibration can support femtosecond pulses to conduct femtosecond laser ablation guided by two-photon florescence imaging at the distal end of their fiber. \cite{Kakkava2019} Tr{\"a}g{\aa}rdh and coworkers  show that a commercial graded-index fiber supports a TM bandwidth sufficient for Coherent Anti-Stokes Raman scattering (CARS) microscopy. \cite{Tragaardh2019} Cifuentes et al. have similarly shown second-harmonic generation imaging through an MMF. \cite{Cifuentes2021} Moving away from applications of ultrathin probes, Leite and coworkers demonstrate far-field imaging of bulk objects with a large depth of field using a large-core step-index MMF. \cite{Leite2021}

An outstanding area of research for MMF endoscopes is on reducing the bend sensitivity of the TM measurement and/or updating the TM calibration from proximal-only signals while remaining confident in the accuracy of the distal control. There have been a few numerical-only proposals to address these limitations, including the use of specialty algorithms with a non-reciprocal reflector \cite{Gu2015} and the use of a stack of metasurface reflectors at three separate wavelengths \cite{Gordon2019,Zheng2023} to enable proximal-only control of the distal electric field. On the experimental side, TMs have been mathematically updated under simple fiber deformation by calculating the fiber shape from a photograph of the fiber to enable continued use of the initial calibration. \cite{Ploschner2015} A distal guidestar has been shown to enable proximal-only imaging over a limited portion of the core of an MMF endoscope; \cite{Li2021} this work, and the work presented in Reference \onlinecite{Gutierrez2023memory}, also highlight the usefulness of the rotational memory effect to obtain control of the distal light fields. Also, it has been proposed that a perfect graded-index profile would greatly simplify bend sensitivity, \cite{Flaes2018} highlighting the importance of fiber design and manufacturing to these applications.

The majority of the MMFs used in these and other related experiments are step- or graded-index fibers, which have been studied under various conditions to target different applications. A few specialty fibers have been studied, including work by Du and coworkers to calibrate a multimode-multicore fiber so that the multimode waveguide can be calibrated for imaging and the single mode cores can be used for sparser imaging and distal light control that is robust to bending. \cite{Du2022} Targeting a different application space, two groups study the transmission matrices of ytterbium-doped MMFs, demonstrating that the control of the spatial profile in such fibers can be used to tailor amplifier or laser performance. \cite{Florentin2019,Sperber2020} Applying the transmission matrix tools to a further array of specialty fibers would enable the benefits of modal control to impact the expanded application space of specialty optical fiber.

As these examples of imaging through MMF endoscopes and on-going research into increasing the robustness of these devices to real-world conditions demonstrate, there is significant interest in and demand for TM measurement setups. To promote this, we present a thorough description of the experimental setup and supporting algorithms to measure and use a polarization-resolved TM, describing our novel way of introducing polarization diversity to the experiment. We utilize an SLM for the basis function generation during the TM measurement and for the subsequent hologram multiplexing, as well as phase-shifting interferometry for the electric field measurement, a Levenberg-Marquardt nonlinear fitting algorithm for the phase and amplitude drift determination that is new in this application, and phase-to-amplitude shaping for the distal image creation. We also present examples demonstrating the reasoning behind our design choices and highlight areas of especial importance in achieving high-quality TM measurements. We emphasize our flexible platform for measuring both scalar and full-polarization TMs, the specific implementation of our phase drift compensation, and present results comparing different phase-to-amplitude multiplexing methods that indicate that other experimental factors matter more to the quality of achievable distal control than the particular choice of a multiplexing method. Throughout this work, we demonstrate our techniques using a variety of specialty optical fibers, including fibers with harsh environment coatings, pedestal fibers, coreless fiber, rectangular core fiber, and a pump-signal combiner, to demonstrate the applicability of these techniques to different fibers and the application spaces they support. 

\section{Overview}
\label{sect:overview}

\begin{figure*}
\centerline{\includegraphics[width=6in]{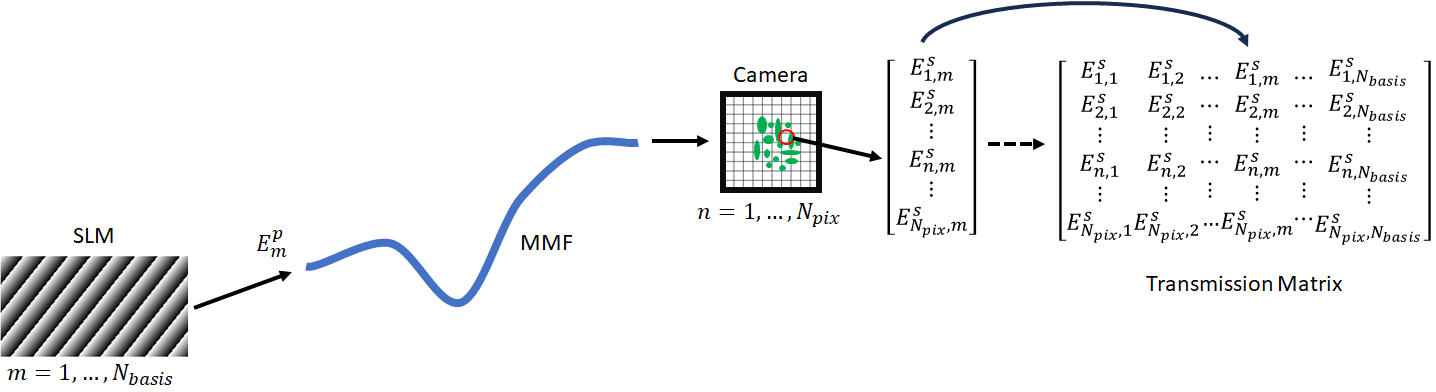}}
\caption{Conceptual overview of the transmission matrix measurement. }
\label{fig:TMOverview}
\end{figure*}

We begin by presenting a conceptual overview of the TM measurement and usage prior to describing our specific implementation in detail. The measurement conceptualization is shown in Fig.\,\ref{fig:TMOverview}. On a phase control device on the proximal end of the fiber, we create a number $N_{\mathrm{basis}}$ of distinct launch basis functions ("holograms") $b_{m}(x,y)$, e.g., linear phase gratings with varying slopes and orientations. It is important to note that $b_{m}$ is a defined input quantity that does not need to be measured. Each of these basis functions $b_{m}$ gives rise to a proximal electric field distribution $E_{m}^{\mathrm{p}}(x,y)$ at the fiber input end.  After propagating through the MMF, each input electric field $E_{m}^{\mathrm{p}}(x,y)$ generates at a camera on the distal side of the fiber an output electric field $E_{m}^{\mathrm{s}}(x,y)$, which is usually a speckle pattern. Putting the $N_{\mathrm{pix}}$ pixel values of $E_{m}^{\mathrm{s}}$ in a column vector, and then stacking all these $N_{\mathrm{basis}}$ column vectors next to each other, we obtain the $N_{\mathrm{pix}}\times N_{\mathrm{basis}}$ transmission matrix $T$. Since each of its columns contains the distal measurement data from one proximal launch condition, the matrix $T$ relates the distal spatial information to the number $m$ of the corresponding proximal launch condition without needing to reveal its proximal transverse distributions $E_{m}^{\mathrm{p}}(x,y)$ and $b_{m}(x,y)$.

To achieve a specific distal electric field $E^{\mathrm{out}}$ (being an $N_{\mathrm{pix}}$-dimensional column vector that contains the intended pixel-by-pixel values of the complex-valued electric field), the TM can be used to solve for an $N_{\mathrm{basis}}$-dimensional launch coefficient column vector $c^{\mathrm{launch}}$ that satisfies the linear system $E^{\mathrm{out}} = T c^{\mathrm{launch}}$ in Eq.\,(\ref{eq:TM}). Hence, there is enormous independence in selecting these basis functions $b_{m}$ as long as the output fields $E_{m}^{\mathrm{s}}$ and thus the columns of $T$ are linearly independent, although some extra consideration is necessary at the top level of performance. \cite{Gomes2022} The computed coefficient vector $c^{\mathrm{launch}}$ contains the real and imaginary part (or amplitude and phase) of the weights that each of the $N_{\mathrm{basis}}$ basis functions at the proximal end needs to be multiplied by so that they coherently sum to produce the desired field $E^{\mathrm{out}}$ at the distal end of the fiber. Using these weights, the basis functions can then be added (or \enquote{multiplexed}) to experimentally generate the intended distal electric field $E^{\mathrm{out}}$. 

If the experimental equipment is a phase-only device such as an SLM or DMD, then only the phase but not the amplitude of the launch conditions (basis functions $b_{m}$ and their superpositions) can be spatially varied. Hence, a computational multiplexing step is required that maps amplitude variations to additional phase variations of the superposition to generate the desired proximal electric field. In essence, that is done by finding a launch phase distribution $\phi^{\mathrm{launch}}(x,y)$ that approximates with the power efficiency constant $c_{\mathrm{eff}}>0$ the superposition in Eq.\,(\ref{eq:multiplexing}).
There are multiple ways to find approximate solutions to Eq.\,(\ref{eq:multiplexing}) with tradeoffs between accuracy and power efficiency, and we test a variety of them and describe the results in Sect. \ref{sec:farFieldResults}. We find that other factors influence the fidelity of the distal control more so than the selection of the multiplexing method.

In summary, the measurement and usage of the transmission matrix consists of three main steps: (1) the creation and launch of a series of linearly independent proximal electric fields, (2) the accurate measurement of the distal electric field created by each proximal electric field, and (3) a multiplexing algorithm to sum the proximal electric fields to generate an arbitrary distal electric field.

To include polarization information in the transmission matrix, two orthogonal input polarizations need to be independently launched at the proximal side of the fiber and measured at the distal side. This effectively doubles the dimension of the launch coefficient vector $c^{\mathrm{launch}}$ to $2N_{\mathrm{basis}}$ and the dimension of $E^{\mathrm{out}}$ to $2N_{\mathrm{pix}}$, thus quadrupling the dimension of the transmission matrix to $2N_{\mathrm{pix}}\times2N_{\mathrm{basis}}$ without changing any of the above-mentioned three main steps.

\section{Experimental Design}
\label{sect:experimentalDesign}

\begin{figure*}
\centerline{\includegraphics[width=6in]{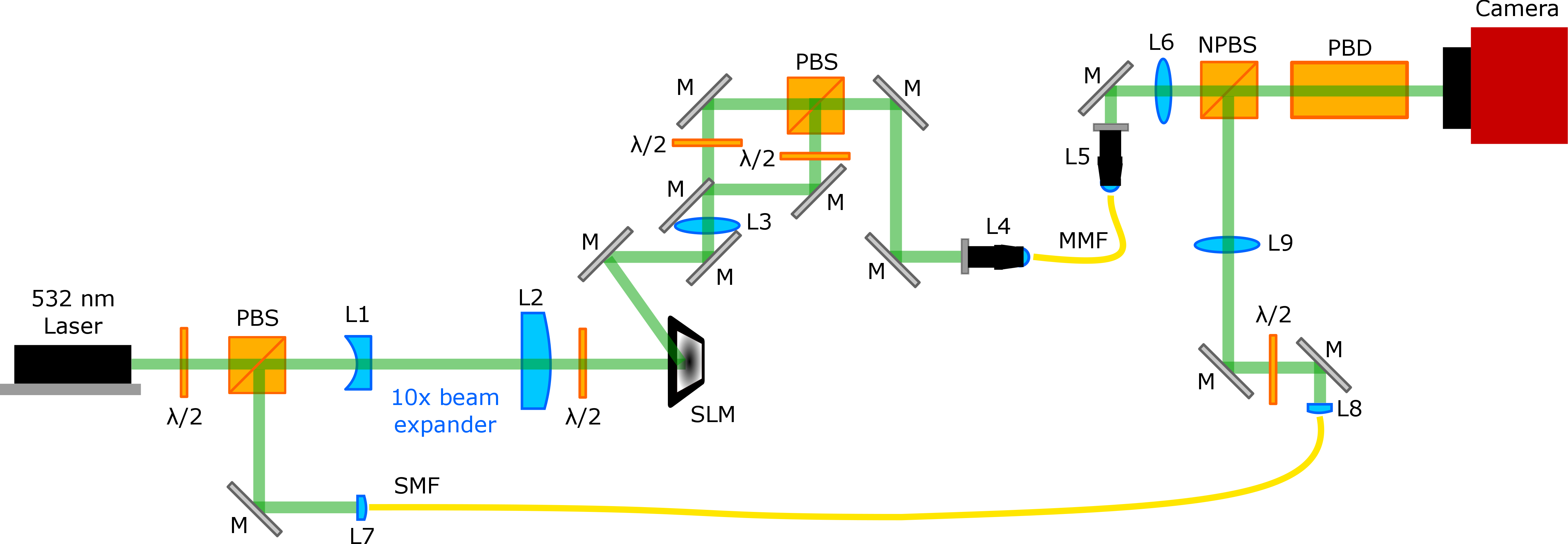}}
\caption{Experimental setup for conducting polarization resolved transmission matrix measurements. $\lambda$/2: half-wave plate; PBS: polarizing beam splitter; L: lens; SLM: spatial light modulator; M: mirror; MMF: multimode fiber; NPBS: non-polarizing beam splitter; PBD: polarizing beam displacer; SMF: single mode fiber. The lenses used are as follows: L1: -20\,mm achromat; L2: 200\,mm achromat; L3: 400\,mm achromat; L4/L5: 20x microscope objectives; L6: 150\,mm achromat; L7/L8: 11\,mm asphere; and L9: 200\,mm achromat.}
\label{fig:Experiment}
\end{figure*}

The experimental setup utilized to measure the polarization-resolved transmission matrix is shown in Fig.\,\ref{fig:Experiment}. We use a linearly polarized continuous wave (CW) laser operating at 532\,nm with an output power of 100\,mW (CrystaLaser CL532-100-S). We place the laser on a thermoelectric (TEC) cooling plate, which we found to reduce long-term drift and noise in the experiments in our laboratory, which experiences seasonal temperature differences. Directly after the laser output, a half-wave plate and a polarizing beam splitter (PBS) serve to split the laser light into two beam paths: one is the signal path that will be coupled to the MMF under study, and the second will provide a reference beam for the interferometric electric field measurement described in Sect. \,\ref{sect:EfieldMeasurement}. 

Following the signal path, the beam next passes through a half-wave plate to match the polarization of the laser light to the preferred polarization of the SLM and a 10x beam expander, formed by lenses L1 and L2. The purpose of the beam expander is to set the beam size slightly less than the size of the SLM (Meadowlark Optics HSP1920-532) screen (under-filling). Although the beam size can also be set to be larger than the SLM screen (over-filling) to maximize the available resolution, we find that under-filling the screen improves the TM measurement due to reflections of unmodulated light from the perimeter of the SLM screen when over-filling. Because of the weight and bulk of the SLM, it is also important to have the SLM mounted in a sturdy, high-precision mount to reduce excess noise in the TM measurement. 

After the SLM, the beam is imaged to the proximal end of the MMF under study through the use of a $4f$ imaging system, \cite{Mertz2010} formed by lenses L3 and the microscope objective L4. Between L3 and L4, the beam path is spatially separated in order to create launch conditions in both orthogonal linear polarizations since the SLM only modulates light in a single linear polarization. A few prior results measure the full polarization-resolved TM, including a demonstration using spatially separated mirrors to generate the two beam paths, \cite{Cizmar2011} and another using a transmissive SLM for full spatial-temporal fiber characterization. \cite{Mounaix2019} To introduce the second polarization, we present a variation on these setups that is flexible and easy to align. We generate two polarized sets of electric field distribution $E_{m}^{\mathrm{p}}(x,y)$ at the proximal fiber end by launching two sets of linear phase gratings that are identical except for the slope of the grating in the x-direction being positive for one set and negative for the other. This allows us to split the basis functions into two different beam paths with the pick-off mirror following L3 and to change the polarization in one of the paths. The beam paths then recombine at the PBS. A half-wave plate is necessary in one of the beam paths in order to switch the polarization of the light modulated by the SLM; one is included in the second beam path as well to keep the beam paths balanced. Although not pictured here, depending on the alignment of the mirrors and other optics, it may be necessary to include irises to block stray beams, such as the higher-diffraction orders from the SLM, from reaching the proximal end of the fiber. The software that runs our experiment is set up to allow for either or both proximal polarizations to be used in a TM measurement through simply indicating which is desired in the program initialization. 

The laser light is next coupled into the MMF, both ends of which are mounted on high-precision three-axis stages (ThorLabs 3-Axis NanoMax Flexture Stages) to enable high-precision coupling. We generally use 20x microscope objectives with numerical apertures of 0.4 at both the proximal and distal ends of the fiber. The alignment of the proximal end of the MMF to either the image or Fourier plane of the SLM (depending on the choice of basis functions) is important. After the initial rough alignment done by measuring the distance between the lenses, we usually optimize this alignment by launching all of the basis functions into the fiber simultaneously and maximizing the output intensity while keeping them centered on the fiber core. Fine tuning after this step is performed by iterating with measurements of the TM itself to minimize its (generalized) condition number, which is the ratio of its largest singular value to its smallest.\cite{Golub}

At the distal end of the fiber, the light emerging from the MMF is imaged to a camera (Photonfocus MV1-D2048x1088-96-G2-10) using the $4f$ imaging system formed by the microscope objective L5 and lens L6. Prior to the camera, a non-polarizing beam splitter (NPBS) is used to combine the signal light with the reference light, and a polarizing beam displacer (PBD) is used to horizontally separate the two linear polarizations. 

The reference light, separated from the signal beam path directly after the laser as stated above, is transmitted closer to the camera using fiber that is single-moded at 532\,nm (SMF). The approximately Gaussian beam profile output from this fiber is also imaged to the camera and overlapped with the signal light using the NPBS to perform the interferometric electric field measurement described in Sect. \,\ref{sect:EfieldMeasurement}. One critical alignment aspect for achieving quality TM measurements is minimizing the tilt between the signal and reference beams at the distal camera. We find that such a tilt, which introduces high-frequency fringes to the interference pattern between the two beams, significantly degrades the accuracy of the distal control that can be achieved with that TM measurement, likely through a combination of accuracy of the electric field measurement and/or sensitivity to noise. 

\section{Basis Functions}
\label{sect:basisFunctions}

In order to measure the transmission matrix, a set of $N_{\mathrm{basis}}$ distinct launch basis functions $b_{m}(x,y)$ is created on a phase control device (e.g., a DMD or SLM as in Figs.\,\ref{fig:TMOverview} and~\ref{fig:Experiment}), each of which generates a complex-valued transverse electric field $E_{m}^{\mathrm{p}}(x,y)$ at the proximal end of the fiber. After propagating to the distal end of the fiber, it generates an output electric field $E_{m}^{\mathrm{s}}(x,y)$ that is also complex-valued and usually describes a speckle pattern. Using the camera on the distal fiber end, $E_{m}^{\mathrm{s}}$ is sampled at $N_{\mathrm{pix}}$ pixels, resulting in an $N_{\mathrm{pix}}$-dimensional column vector. Stacking all these $N_{\mathrm{basis}}$ column vectors next to each other, we obtain the $N_{\mathrm{pix}}\times N_{\mathrm{basis}}$ transmission matrix $T$
as shown in Fig.\,\ref{fig:TMOverview} and described in more detail in Sect. \ref{sect:transmissionMatrix}. Note that the dimension of the transmission matrix is independent of the number of pixels of the SLM.

The proximal end of the fiber can be in either the image \cite{Rothe2019} or Fourier plane \cite{Bianchi2012,Ploschner2014} of the SLM such that the produced input image couples to the MMF core for most basis functions. This is done so that the fiber core can be centered on the set of basis functions by monitoring the transmission of each basis function at the distal end. Successful demonstrations have used discrete portions of the SLM screen \cite{Cizmar2011} and fiber modes. \cite{Carpenter2016,Gu2018} We choose to work with phase gratings to maximize diffraction efficiency from the SLM. \cite{Rosales2017} When using a DMD to modulate the laser light, the Hadamard basis set, \cite{Ohayon2018} Lee holograms, \cite{Lee1979,Zhao2021} and superpixels\cite{Goorden2014} are popular choices.

The SLM used to display the basis functions consists of a nematic liquid crystal screen containing 1920$\times$1152 pixels. By changing the voltage applied to the back-plane of the SLM, the phase imparted to the laser light impinging on that pixel can be changed. Although the voltage-to-phase mapping is not inherently linear, most commercial SLMs come with a look-up table that, when applied, performs this linearization to convert a grayscale image (typically covering 0 to 255 for 8-bit grayscale depth) to the full phase range \cite{Rosales2017} of 0 to $2\pi$.

\begin{figure}
\centerline{\includegraphics[width=3.5in]{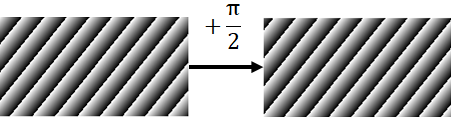}}
\caption{Example of the style of linear phase gratings displayed on the SLM and used to measure the transmission matrix. The period of the grating shown here is larger than in practice so that the detail is visible.}
\label{fig:GratingShift}
\end{figure}

The electric field patterns we chose for our basis functions are generated by imposing linear phase gratings on the SLM. A linear phase grating (Fig.\,\ref{fig:GratingShift}) is a periodic saw-tooth pattern consisting of linear phase slopes covering the full phase range of 0 to $2\pi$ followed by a discontinuous drop back to 0\,rad, which maximizes the first-order diffraction efficiency of the SLM. \cite{Rosales2017} For the lens trains and fibers shown here, each grating period covers approximately 8~pixels. In order to measure both the real and imaginary part (or amplitude and phase) of the resulting electric field at the distal camera, it is necessary to perform at least two independent real-valued measurements of the intensity at the distal camera. This is done by taking the hologram for the launch condition and adding a constant to the grayscale values of the SLM pattern and taking the sum modulo 256 (corresponding to $2\pi$); i.e., effectively adding a constant phase offset to the current basis function. In the case of a linear phase grating as shown in Fig.\,\ref{fig:GratingShift}, this has the effect of looking like the grating has shifted on the SLM screen.  

The linear phase gratings can be used in two different ways to form a useful set of basis functions. In the first, the proximal end of the fiber is placed in the Fourier plane of the SLM, which results in the formation of a spot-shaped electric field $E_{m}^{\mathrm{p}}(x,y)$ at the proximal end of the fiber. By changing the slope of the grating in the x- and y-directions, the location of the spot will change in the Fourier plane, allowing it to be scanned across the input facet of the fiber. Except for minor truncation effects at the transverse boundaries, the transverse Fourier transform that is implicitly performed by the lens train is an orthogonal transformation. Hence, the linear independence of the original gratings on the SLM implies the linear independence of the resulting set of launch conditions at the proximal end of the fiber, which then implies the required linear independence of the distal electric fields $E_{m}^{\mathrm{s}}(x,y)$ if all $E_{m}^{\mathrm{p}}(x,y)$ can be expanded in terms of guided eigenmodes of the MMF. Alternatively, the proximal end of the fiber can be placed in an image plane of the SLM, in which case the same linear phase gratings will produce angled plane waves that also form a linearly independent set $E_{m}^{\mathrm{p}}(x,y)$ at the proximal end of the fiber. Since at least two lenses are used to transmit the laser light from the SLM to the fiber input facet, the beam path in this case passes a Fourier plane which allows for the zero-order reflection and higher diffraction orders to be filtered out from the intended first-order diffraction, but over 90\% of the light remains in the first diffraction order of these phase gratings. \cite{Rosales2017} In our setup, we achieve similar results using either the scanned spot in the Fourier plane or the angled plane waves in the image plane to measure the TM.

The number of basis functions required to adequately measure the transmission matrix depends on the number of modes supported by the fiber, and it is expected that the number of basis functions should be of the same order as the number of fiber modes. \cite{Bianchi2012} In order to ensure the highest signal-to-noise ratio of the distal images created with the measured transmission matrix, it may be advantageous to over-sample the measurement by using a number of launch conditions a few times greater than the number of modes. However, we find that the transmission matrix can be under-sampled by up to a factor of 78 (shown in Fig.\,\ref{fig:200HCS}) and still produce decent quality distal spots. Similarly, in that same experiment, we achieve decent quality distal spots despite using a factor of 118 more $N_\mathrm{pix}$ camera pixels than $N_\mathrm{basis}$ launch conditions. The trade off between measurement time and fidelity of the distal images will be application-dependent.  

\section{Electric Field Measurements}
\label{sect:EfieldMeasurement}

In order to construct the transmission matrix $T$ of the MMF, the transverse distribution of the distal electric field $E_{m}^{\mathrm{s}}(x,y)$ resulting from the launch of each basis function $b_{m}$ needs to be measured. We have chosen to do this via an interferometer with an independent reference beam in a technique known as phase-shifting interferometry. \cite{Cizmar2011} It is also possible to measure the electric field using a co-propagating reference beam, \cite{Popoff2010,Jakl2022} which has the advantage of reducing phase drift in the experiment at the cost of having dark spots in the calibration region or needing to repeat the measurement with multiple choices for the reference beam. Holographic methods, particularly off-axis holography, provide another method of achieving the electric field measurement by interfering the signal beam with a known reference at an angle to the signal beam and considering the Fourier components of the interferometric measurement. \cite{Papadopoulos2012,Dardikman2019} It has the advantage of requiring only one camera frame to measure the electric field but can require additional care with the alignment and is subject to inevitable inaccuracies due to the required spatial frequency filtering that is used to separate the signal and interference components.  

\begin{figure}
    \centerline{\includegraphics[width=3.5in]{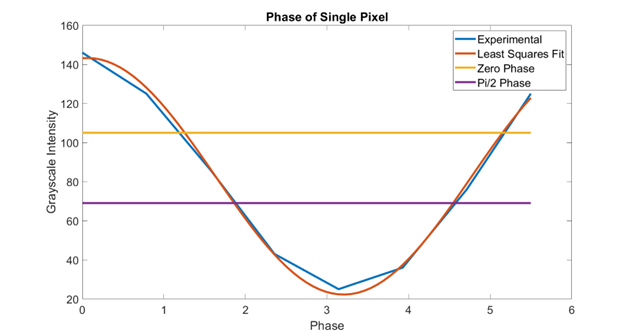}}
\caption{Example of interferometric phase measurement for an individual pixel. Only 3 phase increments are necessary to fit the sinusoidal curve, but 8 were used in this example for visual clarity. The intersection with the two straight lines, representing the intensity at that pixel at a later time with a given additive phase, can be used to determine the amount of phase drift from the initial measurement as described in the main text.}
\label{fig:PhaseMeasurement}
\end{figure}

Our experimental setup is shown in Fig.\,\ref{fig:Experiment}. As described in Sect.\,\ref{sect:experimentalDesign}, a beamsplitter is used to separate the laser beam into two paths directly after its output. One of the beams is coupled into a length of SMF, and the output of the SMF is imaged to the camera. This forms a beam with an approximately Gaussian transverse profile that acts as the reference, and its launch phase is left constant during the measurement. The second beam is reflected from the SLM and coupled into the MMF. The output from the MMF is also imaged to the camera and overlapped with the reference beam. In order to measure the spatial phase of the resulting speckle pattern, a transversely constant phase is added to the phase of the basis function that is launched into the MMF and altered from 0 to $2\pi$ in equidistant steps. According to the Nyquist-Shannon sampling theorem, any number of such equidistant phase steps larger than two (i.e., three or more) is sufficient for the measurement; more may be useful in the presence of noisy data. The electric field is then measured on a pixel-by-pixel basis by constructing the sinusoidal interference curve generated at each camera pixel by the interference of these two beams at said three or more phases. In particular, at the $n$-th distal camera pixel while launching the $m$-th basis function, the intensity $I_{n,m}$ of this interference between the signal beam electric field $E^\mathrm{s}_{n,m}=|E^\mathrm{s}_{n,m}|\eu^{\iu\phi^\mathrm{s}_{n,m}}$ and the reference beam electric field $E^\mathrm{r}_{n,m}=|E^\mathrm{r}_{n,m}|\eu^{\iu\phi^\mathrm{r}_{n,m}}$ is
\begin{align}
    I_{n,m} &= |E^\mathrm{s}_{n,m}+E^\mathrm{r}_{n,m}|^{2}\nonumber\\
    &= \underbrace{|E^\mathrm{s}_{n,m}|^2+|E^\mathrm{r}_{n,m}|^2}_{I^{\mathrm{offset}}_{n,m}}+\underbrace{2|E^\mathrm{r}_{n,m}||E^\mathrm{s}_{n,m}|\cos(\phi^\mathrm{s}_{n,m}-\phi^\mathrm{r}_{n,m})}_{I^{\mathrm{osc}}_{n,m}\cos(\phi^\mathrm{osc}_{n,m})},
    \label{eq:Iinterference}
\end{align}
where $\phi^\mathrm{r}_{n,m}$ and $\phi^\mathrm{s}_{n,m}$ are their corresponding phases. \cite{Heald2012} After determining the three fitting parameters $I^{\mathrm{offset}}_{n,m}$, $I^{\mathrm{osc}}_{n,m}$ and $\phi^\mathrm{osc}_{n,m}$ by fitting the sinusoidal function $I^{\mathrm{offset}}_{n,m}+I^{\mathrm{osc}}_{n,m}\cos(\phi^\mathrm{osc}_{n,m})$ as indicated in Eq.\,(\ref{eq:Iinterference}) to the above-mentioned three or more measurement points that we have for the pixel intensity $I_{n,m}$, we can compute the signal beam electric field in various ways:
\begin{align}
    &E^\mathrm{s}_{n,m}\eu^{-\iu\phi^\mathrm{r}_{n,m}} = \sqrt{I^{\mathrm{offset}}_{n,m}-|E^\mathrm{r}_{n,m}|^2}\eu^{\iu\phi^\mathrm{osc}_{n,m}}\label{eq:Esignal_1}\\
    &= \frac{I^{\mathrm{osc}}_{n,m}}{2|E^\mathrm{r}_{n,m}|}\eu^{\iu\phi^\mathrm{osc}_{n,m}}\label{eq:Esignal_2}\\
    &=\begin{cases}
    \frac{\sqrt{I^{\mathrm{offset}}_{n,m}+I^{\mathrm{osc}}_{n,m}}-\sqrt{I^{\mathrm{offset}}_{n,m}-I^{\mathrm{osc}}_{n,m}}}{2}\eu^{\iu\phi^\mathrm{osc}_{n,m}}\hspace{2.5mm} \text{if } |E^\mathrm{s}_{n,m}|\leq|E^\mathrm{r}_{n,m}|,\\
    \frac{\sqrt{I^{\mathrm{offset}}_{n,m}+I^{\mathrm{osc}}_{n,m}}+\sqrt{I^{\mathrm{offset}}_{n,m}-I^{\mathrm{osc}}_{n,m}}}{2}\eu^{\iu\phi^\mathrm{osc}_{n,m}} \hspace{2.5mm}\text{otherwise.}
    &\end{cases}
    \label{eq:Esignal_3}
\end{align}
While Eqs.\,(\ref{eq:Esignal_1}) to~(\ref{eq:Esignal_3}) are equivalent if $|E^\mathrm{r}_{n,m}>0|$ and in the absence of noise (i.e., vanishing residual of the fit in Eq.\,(\ref{eq:Iinterference})), they have different stability properties in the presence of noise. Equations~(\ref{eq:Esignal_1}) and~(\ref{eq:Esignal_2}) require exact knowledge of the reference beam electric field modulus $|E^\mathrm{r}_{n,m}|$ at the \emph{exact same} time at which $I_{n,m}$ is measured, as well as the conditions $I^{\mathrm{offset}}_{n,m}\ge|E^\mathrm{r}_{n,m}|^2$ and $|E^\mathrm{r}_{n,m}|>0$, respectively. In contrast, Eq.\,(\ref{eq:Esignal_3}) only requires knowing if $|E^\mathrm{s}_{n,m}|\leq|E^\mathrm{r}_{n,m}|$ is true. Since $|E^\mathrm{s}_{n,m}|^2$ is a speckle pattern that can become zero at some or many pixels, it would be impossible to guarantee $|E^\mathrm{s}_{n,m}|>|E^\mathrm{r}_{n,m}|$, but, with a sufficiently strong reference beam, it is easy to guarantee $|E^\mathrm{s}_{n,m}|\leq|E^\mathrm{r}_{n,m}|$. Hence, we use the upper part of Eq.\,(\ref{eq:Esignal_3}) to determine the signal beam electric field $E^\mathrm{s}_{n,m}$ at all pixels $n=1,\ldots,N_{\mathrm{pix}}$.

Since only the relative phase $\phi^\mathrm{s}_{n,m}-\phi^\mathrm{r}_{n,m}$ between the signal and reference beams can be determined using the interferometric Eqs.\,(\ref{eq:Iinterference}) to~(\ref{eq:Esignal_3}), it is preferable to align the experiment such that $\phi^\mathrm{r}_{n,m}$ is the same for all pixels, i.e., independent of $n$, if both the intensity and phase of the desired distal electric field matters. To compensate for fluctuations of the reference phase $\phi^\mathrm{r}_{n,m}$ while we scan through the $m=1,\ldots,N_{\mathrm{basis}}$ basis functions, we use the phase drift compensation technique described in the following Sect.\,\ref{sect:phaseDrift}.
While the fitting procedure (over the at least three signal phases as described above) to determine the electric field, e.g., by Eq.\,(\ref{eq:Esignal_3}), is performed separately for each pixel, the fit to determine the phase drift in Sect.\,\ref{sect:phaseDrift} is performed simultaneously for all pixels.

\section{Measuring Phase Drift}
\label{sect:phaseDrift}

The reference phase $\phi^\mathrm{r}_{n,m}$ drifts during the time it takes to scan all basis functions $b_{m}$ with $m=1,\ldots,N_{\mathrm{basis}}$ in Eq.\,(\ref{eq:Esignal_3}) used to measure the transmission matrix $T$. Since only the relative phase $\phi^\mathrm{s}_{n,m}-\phi^\mathrm{r}_{n,m}$ can be determined by the interferometric measurement technique described by Eq.\,(\ref{eq:Iinterference}), this $m$-dependent drift of $\phi^\mathrm{r}_{n,m}$ needs to be measured and subtracted from the data according to the left hand side of Eq.\,(\ref{eq:Esignal_1}) in order to produce a consistent transmission matrix. Without this procedure, it is impossible to correctly control the distal electric field.

The phase drift is tracked during the data acquisition by first picking a launch condition $m_{\mathrm{ref}}$ to serve as the phase drift reference. We pick one that is near the center spatial frequency of the set of basis functions $b_{m}(x,y)$ in use. It could be one of those basis functions, but we usually choose a reference launch condition that is not identical to any of the basis functions. Prior to collecting any of the data for the main set of basis functions, the interferometric intensity $I_{n,m_{\mathrm{ref}}}$ of that reference launch condition is measured at all pixels $n=1,\ldots,N_{\mathrm{pix}}$ with at least three signal phases as described in Sect.\,\ref{sect:EfieldMeasurement}. Similar to Eq.\,(\ref{eq:Iinterference}), this results in determining the fitting parameters $I^{\mathrm{offset}}_{n,m_{\mathrm{ref}}}$, $I^{\mathrm{osc}}_{n,m_{\mathrm{ref}}}$ and $\phi^\mathrm{osc}_{n,m_{\mathrm{ref}}}$ using the reference fit $I_{n,m_{\mathrm{ref}}}=I^{\mathrm{offset}}_{n,m_{\mathrm{ref}}}+I^{\mathrm{osc}}_{n,m_{\mathrm{ref}}}\cos(\phi^\mathrm{osc}_{n,m_{\mathrm{ref}}})$. We usually repeat this procedure a number of times (e.g., 100) and keep the highest-quality repetition, i.e., the launch that had the smallest residual of its fit. This fit creates a reference at \enquote{time zero} ($t_{0}$). Then, at time $t_{m}$, which needs to be immediately before or after (simultaneously is impossible) measuring the interference intensity $I_{n,m}$ of any regular basis function $m$ at all pixels $n=1,\ldots,N_{\mathrm{pix}}$, the reference launch condition $m_{\mathrm{ref}}$ is re-launched into the fiber to determine the phase $\phi^\mathrm{r}_{n,m_{\mathrm{ref}}}(t_{m})$ as described above and thus the phase drift
\begin{equation}\phi^{\mathrm{drift}}_{m}=\phi^\mathrm{r}_{n,m_{\mathrm{ref}}}(t_{m})-\phi^\mathrm{r}_{n,m_{\mathrm{ref}}}(t_{0})
\approx\phi^\mathrm{r}_{n,m}-\phi^\mathrm{r}_{n,m_{\mathrm{ref}}}(t_{0}).
\label{eq:phaseDrift}
\end{equation}
The validity of the approximation in Eq.\,(\ref{eq:phaseDrift}) requires that $I_{n,m_{\mathrm{ref}}}$ is measured immediately before or after $I_{n,m}$ such that their phase drifts can be assumed to be approximately identical.

To evaluate $\phi^{\mathrm{drift}}_{m}$ on a pixel-by-pixel basis, we would need to relaunch the reference $m_{\mathrm{ref}}$ with at least two different additive phase increments. As an example, Fig.\,\ref{fig:PhaseMeasurement} shows the two horizontal lines, which are the intensity values $I_{n,m_{\mathrm{ref}}}$ of the same pixel $n$ from the re-launch of the reference launch condition $m_{\mathrm{ref}}$ with 0-phase (yellow line) and $\pi/2$ phase (purple line) taken subsequently. The intersection of these horizontal lines with the original sinusoidal fitting curve from $m_{\mathrm{ref}}$ in Fig.\,\ref{fig:PhaseMeasurement} tells us how much the phase of the interferometer has changed at pixel $n$ since the initial calibration. Because the intensity line intersects the sine curve twice, in general, two phases have to be checked to determine which phase drift of the two possibilities is correct. However, under our assumption that the phase drift $\phi^{\mathrm{drift}}_{m}$ is independent of the particular pixel $n$, a single relaunch of the reference launch condition $m_{\mathrm{ref}}$ is sufficient by using a fitting process across all pixels. Nevertheless, we continue to record both additive phases $0$ and $\pi/2$ to have the pixel-by-pixel phase drift as well.

\begin{figure*}
    \centerline{\includegraphics[width=7in]{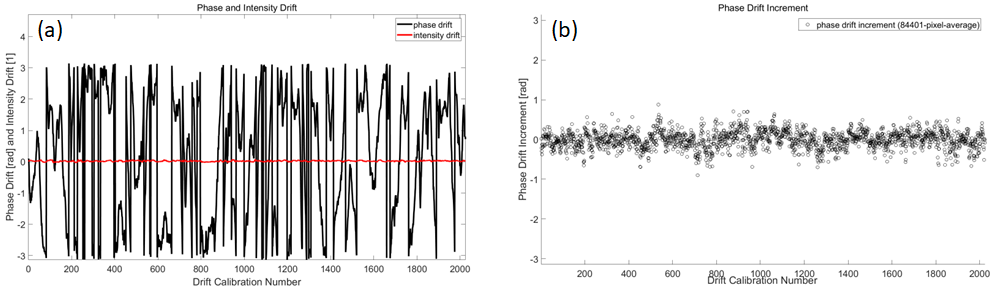}}
\caption{(a) Time series of phase and intensity drift versus basis function number. (b) The increment of the phase drift versus basis function number.}
\label{fig:PhaseDriftCurve}
\end{figure*}

To simultaneously track the phase drift $\phi^{\mathrm{drift}}_{m}$ and a possible amplitude drift $a^{\mathrm{drift}}_m$ while we launch the $m=1,\ldots,N_{\mathrm{basis}}$ basis functions $b_{m}$, we use a Levenberg-Marquardt method \cite{Levenberg1944,Marquardt1963} to solve the following nonlinear optimization problem:
\begin{align}
\begin{bmatrix}\phi^{\mathrm{drift}}_{m}\\a^{\mathrm{drift}}_m\end{bmatrix}=
\argmin_{\scriptsize{\begin{bmatrix}\phi\in[0,2\pi)\\a>0\end{bmatrix}}}
\sum_{n=1}^{N_{\mathrm{pix}}}\biggl|&a\left[I^{\mathrm{offset}}_{n,m_{\mathrm{ref}}}+I^{\mathrm{osc}}_{n,m_{\mathrm{ref}}}\cos(\phi^{\mathrm{osc}}_{n,m_{\mathrm{ref}}}+\phi)\right]\nonumber\\
&-I_{n,m_{\mathrm{ref}}}(t_{m})\Bigr|^2.\label{eq:phaseAndAmplitudeDrift}
\end{align}
The inclusion of an amplitude drift in Eq.\,({\ref{eq:phaseAndAmplitudeDrift}}) increases the accuracy of the drift compensation because it keeps amplitude fluctuations of the speckle pattern $I_{n,m_{\mathrm{ref}}}(t_{m})$ from being misinterpreted as additional phase fluctuations. Nevertheless, a simpler phase drift-only model that implicitly sets $a=1$ in Eq.\,({\ref{eq:phaseAndAmplitudeDrift}}) usually catches the majority of the drift effects.

This process of measuring the drift reference interferogram $I_{n,m_{\mathrm{ref}}}$ is performed immediately before or after collecting the data to measure the electric field for each basis function. From these data, the time series $\phi^{\mathrm{drift}}_{m}$ and $a^{\mathrm{drift}}_m$ mapping the interferometer's phase and amplitude drift during the data acquisition can be created; see Fig.\,\ref{fig:PhaseDriftCurve}, which shows that the relative phase in the interferometer can vary quite dramatically during a transmission matrix measurement. These phase drift values need to be added to the reference phase $\phi^\mathrm{r}_{n,m}$ in the electric field Eq.\,(\ref{eq:Esignal_1}) according to Eq.\,(\ref{eq:phaseDrift})
\begin{equation}\phi^\mathrm{r}_{n,m}\approx\phi^\mathrm{r}_{n,m_{\mathrm{ref}}}(t_{0})+\phi^{\mathrm{drift}}_{m}.
\label{eq:phaseDriftSubtraction}
\end{equation}
This ensures that the spatial phase for each basis function launch is calculated with respect to the same \enquote{time-zero} reference beam, which is necessary to have a consistent framework for the transmission matrix constructed from these data. 

\section{Data Acquisition Process}

The data required to measure the transmission matrix and environmental phase drift consist of thousands to tens of thousands of (typically) 8-bit camera images. We have chosen to use MATLAB to control much of the experiment. It can load a series of pre-saved basis functions $b_{m}$ (holograms) and send them to the SLM, which is connected to the computer via a PCIe port. The SLM displays these images in succession. We typically run the SLM slightly below its maximum refresh rate of 100\,Hz to reduce the incidence of dropped frames, which become virtually nonexistent around 80\,Hz with our equipment. Newer SLMs and DMDs can operate at orders-of-magnitude faster refresh rates. 

\begin{figure*}
    \centerline{\includegraphics[width=6in]{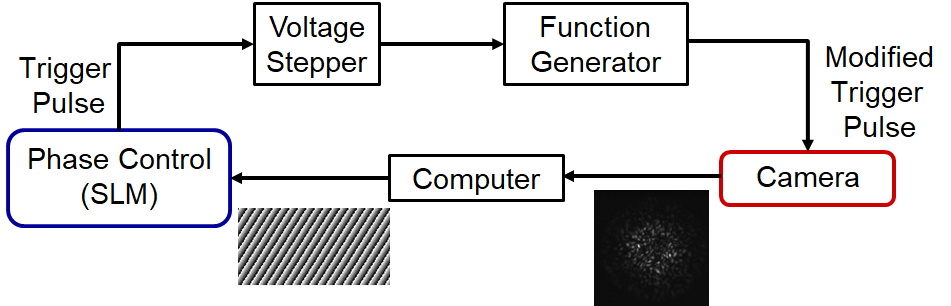}}
\caption{Data collection and triggering process.}
\label{fig:DataAcquisition}
\end{figure*}

Next, the SLM controller produces a trigger pulse when a new image is sent to the SLM. Ideally, this electronic pulse would directly trigger the camera to record the distal image associated with the pattern displayed on the SLM, but, using our specific equipment, the camera cannot detect pulses as short as the one produced by the SLM. Instead, the SLM's trigger pulse is increased in amplitude via a stepper circuit (SparkFun Logic Level Converter BOB-12009), and this pulse is used to trigger a function generator, which in turn creates a pulse that can trigger the camera to record an image. We delay the camera trigger pulse by 0.5-1\,ms to ensure that the liquid crystals on the SLM screen have finished rotating prior to recording the camera image; without a delay, the recorded camera image can contain signal light from both the current and prior SLM holograms. In the other extreme of too long a delay, the phase drift during the measurement increases and the quality of the TM measurement degrades. We empirically determined the optimum delay by monitoring the quality of our TM measurements. The images collected by the camera are then saved to the same computer that is sending the SLM holograms to the experiment. Figure~\ref{fig:DataAcquisition} represents our typical data collection process. 

\begin{figure*}
    \centerline{\includegraphics[width=6in]{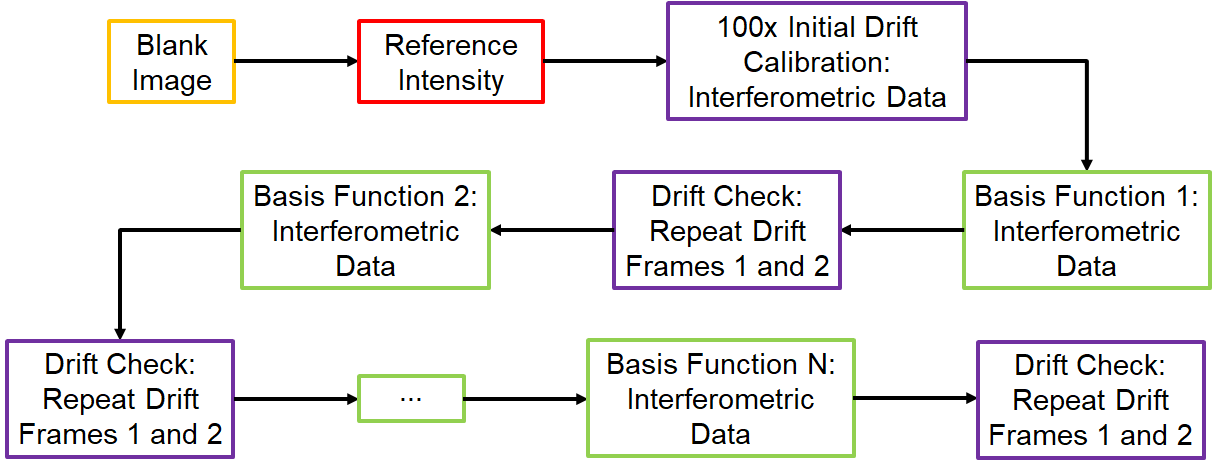}}
\caption{Example of typical data structure.}
\label{fig:DataFlowChart}
\end{figure*}

The order of a typical transmission matrix data set is shown in Fig.\,\ref{fig:DataFlowChart}. We begin by recording a blank camera image so that the camera background can be determined and used for post-processing of the rest of the data. Second, the intensity of the reference beam only, $I_{m_\mathrm{ref}}$, is recorded so that the intensity of the speckle patterns can be reconstructed from the interferometric data as discussed in Sect.\,\ref{sect:phaseDrift}. Third, we record the interferometric data $I_{n,m_{\mathrm{ref}}}$ to determine the phase of the drift calibration launch with respect to the reference beam 100 times. We use the redundant data to find the phase measurement with the minimal residual to the fitted cosine curve in order to start with as accurate as possible information for the phase of the drift calibration launch; this step is not strictly necessary.
Fourth, the main section of transmission matrix data $I_{n,m}$ is recorded. This is done by taking each basis function and recording the (usually 3, which is the required minimum as described in Sect.\,\ref{sect:EfieldMeasurement}) interferometric frames to determine its phase. After each basis function launch, reference launch conditions $m_{\mathrm{ref}}$ with 0 additive phase and $\pi/2$ additive phase are re-displayed to the SLM to determine the phase drift over time as described in Sect.\,\ref{sect:phaseDrift}. This process is repeated until the interferometric data and drift frames are recorded for each of the N basis functions used to measure the transmission matrix. 

It is worth noting that it is critically important to have a reliable data stream and method for collecting and labeling the images in order to construct a useful transmission matrix. For the camera in use for our experiments, we have empirically determined that it records 3 frames upon initialization that are not part of our desired data stream, so these frames are always discarded. We have also found it advantageous, especially during initial troubleshooting of our experiment, to record blank (or reference-only) buffer camera images between the main sections of data in order to delineate the sections of data and ensure that all of the expected frames are being recorded. 

\section{Creating the Transmission Matrix}
\label{sect:transmissionMatrix}

After collecting a data set and accounting for the measured phase drift as described in Sect.\,\ref{sect:phaseDrift}, a few other important pre-processing steps are performed prior to populating the transmission matrix. Firstly, to save memory and computation time, we consider the distal camera images collected for the launch condition $m_{\mathrm{ref}}$ to be used for the environmental drift calculation and crop them and all others so that they only contain pixels $n$ that have both sufficient signal $|E^\mathrm{s}_{n,m_{\mathrm{ref}}}|^2$ and reference light $|E^\mathrm{r}_{n,m_{\mathrm{ref}}}|^2$ in Eq.\,(\ref{eq:Iinterference}). By requiring a certain minimum interferometric oscillation intensity $I^{\mathrm{osc}}_{n,m_{\mathrm{ref}}}$ in Eq.\,(\ref{eq:Iinterference}), we define the set of pixels that we then use for all of our data, and define $N_{\mathrm{pix}}$ as the number of pixels in this set. We also subtract the average camera background from the images, setting negative pixel values to 0. Secondly, we screen the basis functions to make sure only those that produced high-quality data are included in the transmission matrix. After testing multiple metrics for accomplishing this, we chose to discard launch conditions $m$ with power below a certain threshold and launch conditions with a cosine-fit residual in Eq.\,(\ref{eq:Iinterference}) above a certain threshold. The optimal values for these thresholds depend on the specifics of the data set; we typically set them so that about 10\% of the basis functions fail the threshold and are discarded. Note that there can be overlap between the launch conditions flagged by both of these metrics, and that for many data sets these choices have a minor impact on the fidelity of the distal goal images. 

\begin{figure*}
    \centerline{\includegraphics[width=7in]{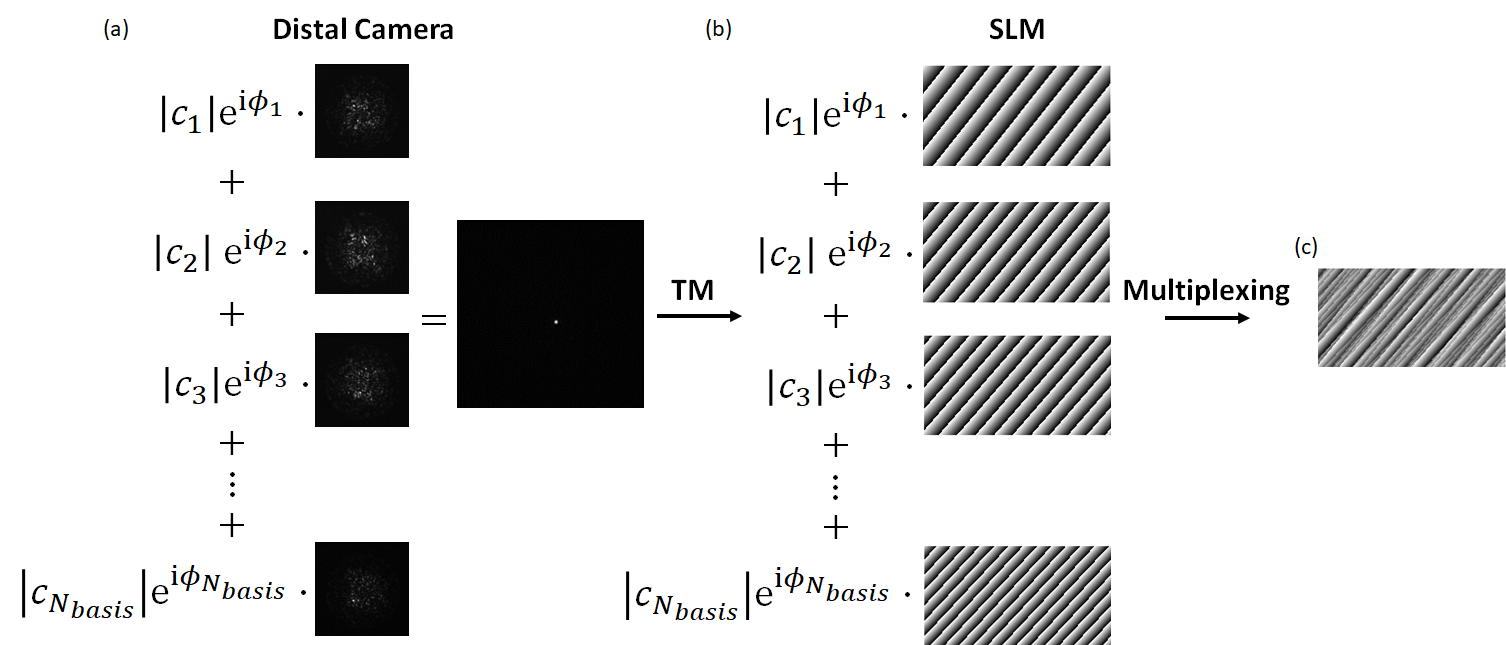}}
\caption{(a) Example of correctly summing the speckle patterns to create the desired distal pattern, in this case a focused spot. (b) Corresponding SLM view showing the transmission matrix predicting the correct complex summation of the basis functions (in this case, linear gratings) that is desired to generate a focused spot. (c) Multiplexing step that transforms the predicted complex superposition of gratings into a phase-only superposition of gratings that can be displayed on the phase-only spatial light modulator (SLM).}
\label{fig:PhaseWeights}
\end{figure*}

Then, the transmission matrix can be constructed from the measured data. Using the symbol $\mathbb{C}$ to denote the set of complex numbers, it is the matrix $T\in\mathbb{C}^{N_\mathrm{pix}\times N_\mathrm{basis}}$ as described in Sect.\,\ref{sect:basisFunctions} that gives the electric field $E^{\mathrm{out}}\in\mathbb{C}^{N_\mathrm{pix}}$ at the distal camera from any given superposition of basis functions, quantified by the launch coefficient vector $c_{\mathrm{launch}}\in\mathbb{C}^{N_\mathrm{basis}}$:
\begin{equation}
    E^{\mathrm{out}} = Tc^{\mathrm{launch}}.
    \label{eq:TM}
\end{equation}
The entries of the input vector $c_{\mathrm{launch}}$ are the weights for the individual basis functions in the superposition that is launched by the SLM.  Similarly, each entry of the output vector $E^{\mathrm{out}}$ quantifies the complex-valued electric field at a single pixel in the distal speckle pattern. Accordingly, each column of the transmission matrix contains the measured complex-valued electric field at all relevant pixels of the distal output camera when a certain basis function is launched, as described in Sect.\,\ref{sect:basisFunctions}. 

We ultimately want to use the measured transmission matrix to create specific distal outputs from the MMF. According to Eq.\,(\ref{eq:TM}) above, we need to solve (e.g., in a least-squares sense in the typical overdetermined case with $N_\mathrm{pix}>N_\mathrm{basis}$) for the input vector $c_{\mathrm{launch}}$ such that the output vector $E^{\mathrm{out}}$ contains a specified target image, such as a single bright pixel at a specific location or an entire image that extends over basically all distal pixels. This can be done using any linear algebra package (in our case, MATLAB).

The result of this calculation is a vector $c_{\mathrm{launch}}$ containing the intensity and phase that each launch condition needs to be multiplied by so that when all launch conditions are produced simultaneously by the SLM in a process called multiplexing, their individual distal speckle patterns add up to produce the desired distal image. This is depicted in Fig.\,\ref{fig:PhaseWeights}(a): when the speckle pattern corresponding to each launch condition is multiplied by the correct factor and summed, a focused spot can be achieved. In Fig.\,\ref{fig:PhaseWeights}(b), we show the corresponding SLM view where the linear phase gratings (the basis functions) are all multiplied by the pre-determined complex phase and amplitude predicted by the TM. Since we are using a phase-only modulation device, one further step is required before the SLM pattern can be created and used. This step takes the predicted complex summation and transforms it into a phase-only superposition in a process called multiplexing. This is shown in Fig.\,\ref{fig:PhaseWeights}(c) and described in Sect.\,\ref{sect:multiplexing} below. 

\section{Multiplexing Holograms}
\label{sect:multiplexing}

\begin{figure*}
    \centerline{\includegraphics[width=6in]{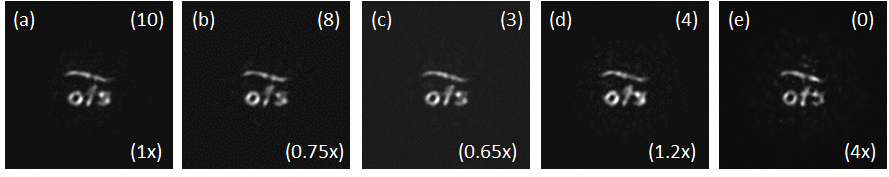}}
\caption{Comparison of various hologram multiplexing methods used to create an image in the image plane of the distal end of the calibrated fiber (OFS TCU-ME050H graded-index MMF) in a single polarization: (a) Reference [\onlinecite{Davis1999}]; (b) Reference [\onlinecite{Bolduc2013}], equation 4; (c) Method 3 in reference [\onlinecite{Arrizon2007}]; (d) Destructive interference between neighboring pixels with 2-pixel groups; \cite{Rosales2017} (e) Add phasors with all amplitudes set to unity. The numbers in the lower left indicate the relative power efficiency of the multiplexing method, referenced to panel (a). Size of images is 825\,µm, which is approximately a 13x magnification of the fiber endface. Fiber calibrated with 45$\times$45 angled plane waves.}
\label{fig:Multiplexing}
\end{figure*}

By definition, a phase-only SLM can only modulate the phase of the light incident upon it. In this and numerous other applications, though, it is desirable to control both the phase and the amplitude. This can be done by using a variety of established phase-to-amplitude shaping techniques. 

The SLM can only decrease the amplitude of the light; so, all of the phase-to-amplitude techniques shape the amplitude by sending light away from the desired location and thus come at a cost to overall efficiency. Assuming that the first-order diffraction from the SLM will contain the desired modulated light, amplitude shaping can be performed by sending light in a spatially-varying fashion from the first order diffraction to either the zero-order reflection by adjusting the depth of the grating (i.e., the grating does not cover the whole range of 0 to 2$\pi$) or to the higher diffraction orders by varying the spatial frequency of the grating. \cite{Rosales2017,Davis1999} These concepts can be applied to non-grating holograms by superimposing the additional desired pattern with a linear phase grating. In this case, the linear phase grating will transmit the pattern to the first diffraction order of the grating, which can be spatially isolated in a Fourier plane of the SLM. \cite{Rosales2017}

There are numerous ways to implement these concepts mathematically in order to achieve phase and amplitude shaping from the SLM with trade-offs between accuracy, efficiency, and sensitivity to alignment. With $c_{m}^{\mathrm{launch}}$ being the entries of the input vector $c_{\mathrm{launch}}$ computed in Eq.\,(\ref{eq:TM}) and $b_{m}$ being the basis functions mentioned in Sect.\,\ref{sect:basisFunctions}, all these methods focus on different ways to achieve the task illustrated in in Fig.\,\ref{fig:PhaseWeights}(c), i.e., finding approximate solutions to the equation 

\begin{equation}    \sqrt{c_{\mathrm{eff}}}\sum_{m=1}^{N_{\mathrm{basis}}}c_{m}^{\mathrm{launch}}b_{m}(x,y)\approx\mathrm{e}^{\mathrm{i}\phi^{\mathrm{launch}}(x,y)}.\label{eq:multiplexing}
\end{equation}

We have tried the multiplexing methods presented in References [\onlinecite{Rosales2017,Davis1999,Arrizon2007,Bolduc2013, Davis1999}]. Figure~\ref{fig:Multiplexing} shows the same goal image created in the image plane of the distal fiber endface, which matches the experimental geometry shown in Fig.\,\ref{fig:Experiment}, via five of these techniques. With our setup, we have found the method in Reference [\onlinecite{Davis1999}] to provide the best balance of these factors in our transmission matrix application with the linear phase grating basis functions (Fig.\,\ref{fig:Multiplexing}, panel (a)). These experiments were carried out in a specialty graded-index fiber with both the core and coating optimized to withstand harsh, high temperature environments where hydrogen darkening is of concern. 

\section{Alternate Fiber Geometries}
\label{sect:alternateFibers}

As discussed in Section \ref{sec:introduction}, the majority of published TM results focus on standard step-index and graded-index MMFs. Here, we highlight distal image creation through a variety of specialty fibers to illustrate the broad applicability of our TM measurement setup to different fibers. Rather than optimizing the number of basis functions (linear phase gratings) and their parameters (slope increments on the SLM, corresponding to spot spacings in the Fourier plane) to each fiber, we show results using the same measurement for each fiber; results could be further optimized by adjusting the basis functions for each one individually. 

\begin{figure}
    \centerline{\includegraphics[width=3.25in]{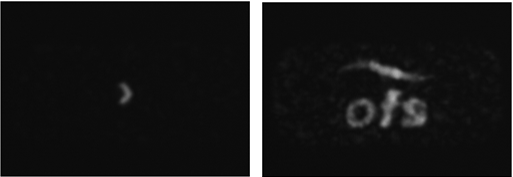}}
\caption{Example distal images created through approximately a meter of rectangular core fiber with core size 43.5\,µm$\times$96.7\,µm. Image sizes are 1$\times$1.375\,mm at approximately 13x magnification. The fiber was calibrated with a grid of 35$\times$35 scanned spots.}
\label{fig:RectangularFiber}
\end{figure}

In Fig.\,\ref{fig:RectangularFiber}, we present distal goal images created through a rectangular core fiber with a core size of 43.5\,µm$\times$96.7\,µm and having a cladding glass diameter of 125\,µm. There are applications where the geometry of the rectangle may better suit the end use of the fiber. However, the modes of this fiber will differ from those of the more standard circular core fibers, and we note that we achieve similar performance from this waveguide using the basis functions and TM measurement optimized for the circular fibers. 

\begin{figure}
    \centerline{\includegraphics[width=3.25in]{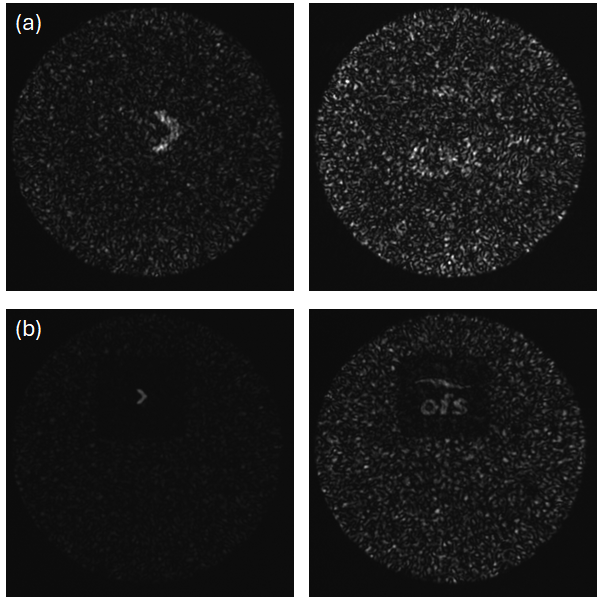}}
\caption{Distal images created through approximately a meter of hard clad silica (HCS) fiber with a core diameter of 200\,µm. (a) The entire core region is calibrated. (b) A restricted region (the black square) is calibrated. Image sizes are 2.75\,mm$\times$2.75\,mm at approximately 13x magnification. The fiber was calibrated with a grid of 35$\times$35 scanned spots.}
\label{fig:200HCS}
\end{figure}

We next calibrate a hard-clad silica (HCS\textsuperscript{\textcopyright}) fiber with a numerical aperture (NA) of 0.37 and a 200\,µm core diameter, which is pure silica, surrounded by a low index polymer. This step-index waveguide supports over 95,000 modes at our 532\,nm wavelength, but we are probing the TM with only $N_{\mathrm{basis}}=1225$ linear phase gratings as basis functions $b_{m}$, with slopes chosen such that they result in a grid of 35$\times$35 spots at the proximal fiber end, which in this case is in the Fourier plane of the SLM. Given this massive undersampling ratio of approximately 78, it is remarkable that in Fig.\,\ref{fig:200HCS}(a), we are still able to achieve faintly visible extended images at the distal end of the fiber. In fact, the distal spots (not pictured) have a sufficiently high signal-to-noise ratio (SNR) to achieve USAF target images, as shown in Section \ref{sec:farFieldResults}, Fig. \ref{fig:USAFTargets}(a). The quality of the extended images shown here can be improved by increasing the number $N_{\mathrm{basis}}$ of basis functions used in the measurement; instead, in Fig \ref{fig:200HCS}(b), we restrict the calibration region to the black square in the upper half of the fiber core and show the improvement in the image quality over the smaller area. This also illustrates the contrast between the controlled black square and the remaining speckle pattern in the non-calibrated region. This style of calibrating the fiber is potentially useful to applications such as dark field imaging. \cite{Eugui2018}

\begin{figure}
    \centerline{\includegraphics[width=3.25in]{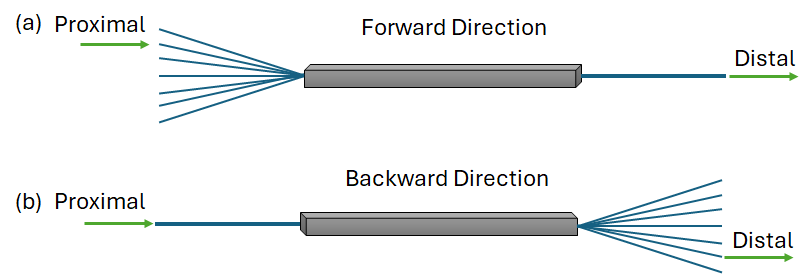}}
\caption{(a) Schematic of the pump-signal combiner operating in the forwards direction. (b) Schematic of the pump-signal combiner operating in the backwards direction.}
\label{fig:CombinerSchematics}
\end{figure}

\begin{figure}
    \centerline{\includegraphics[width=3.25in]{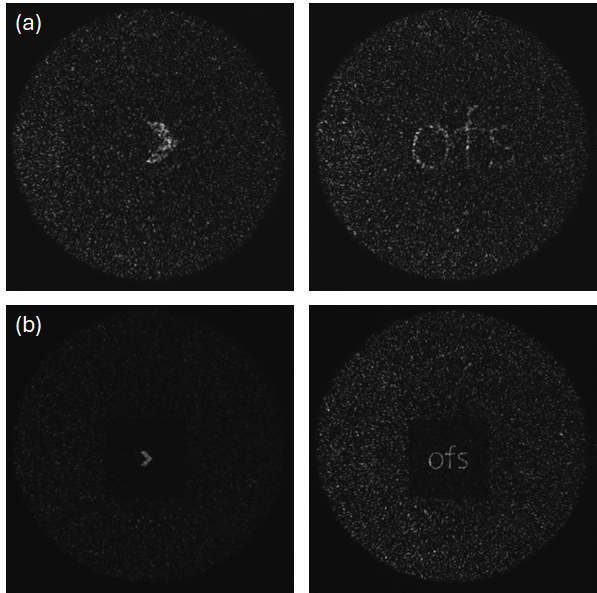}}
\caption{Distal images created through a 7$\times$1 pump-signal combiner by launching light into one of the 110/125\,µm pump leads and placing the output of the 220/240\,µm at the distal end (a) The entire core region is calibrated. (b) A restricted region (the black square) is calibrated. Image sizes are 3x3\,mm at approximately 13x magnification. The fiber was calibrated with a grid of 35$\times$35 scanned spots.}
\label{fig:PSCForward}
\end{figure}

\begin{figure}
    \centerline{\includegraphics[width=3.25in]{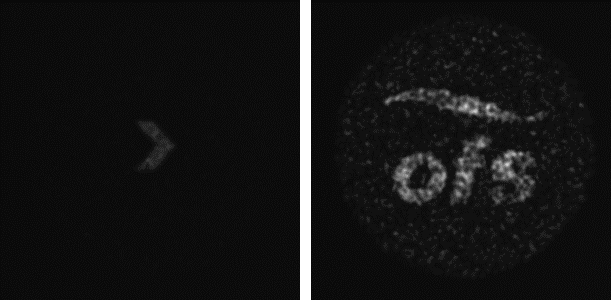}}
\caption{Distal images created through a 7$\times$1 pump-signal combiner by launching light into the 220/240\,µm output fiber and calibrating the output at one of the 110/125\,µm pump lead fibers. Image sizes are 1.65x1.65\,mm at approximately 13x magnification. The fiber was calibrated with a grid of 35$\times$35 scanned spots.}
\label{fig:PSCBackward}
\end{figure}

In a final example, we measure the TM of a 7$\times$1 pump-signal combiner (PSC) based on a tapered fiber bundle (TFB) operating in both directions (i.e., illuminating a single pump input and measuring the TM at the output fiber and vice-versa) as shown in Fig.\,\ref{fig:CombinerSchematics}. In Fig.\,\ref{fig:PSCForward}, we show the distal goal images for the forward direction, showing a similar trade-off in imaging quality to calibration region as with the 200\,µm HCS fiber in Fig.\,\ref{fig:200HCS}. The results for the backward direction are shown in Fig.\,\ref{fig:PSCBackward}. This illustrates a couple of points. First, few TM measurements have been of systems that contain a non-homogeneous transverse profile, as exists here due to the taper. Second, we have a mismatch in he mode content of the input and output fibers, which are 110/125\,µm and 220/240\,µm, respectively, and are able to control the distal field well in both directions. Thirdly, when operating the PSC in the forward direction, it is sufficient to launch light into a random one of the pump leads and still be able to calibrate the entire core of the output fiber, although there is some light enhancement in the region of the output fiber closest to that input. Finally, other groups have studied the TMs of ytterbium-doped MMFs, demonstrating that the control of the spatial profile in such fibers could be use to tailor amplifier or laser performance, \cite{Florentin2019,Sperber2020} but they relied on free-space coupling into the gain fiber. Utilizing a PSC in these applications would allow them to leverage the benefits of having an all-fiber system, and this demonstrates that its TM can be measured in conjunction with that of the gain fiber. 

\section{Full Polarization Results}

\begin{figure*}
    \centerline{\includegraphics[width=4in]{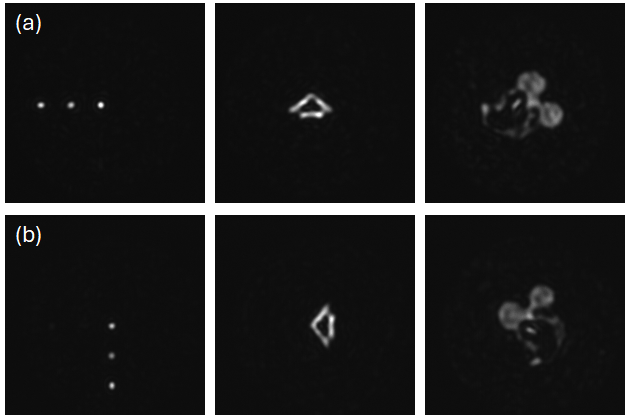}}
\caption{A set of focused spots, a company logo, and a character created in a pedestal-style fiber in the (a) horizontal polarization and (b) vertical polarization created with the full vector transmission matrix. Size of images is 1.1\,mm, which is approximately a 13x magnification of the fiber endface. Fiber calibrated with 35$\times$35 scanned spots.}
\label{fig:PolarizationLogos}
\end{figure*}

\begin{figure*}
    \centerline{\includegraphics[width=4in]{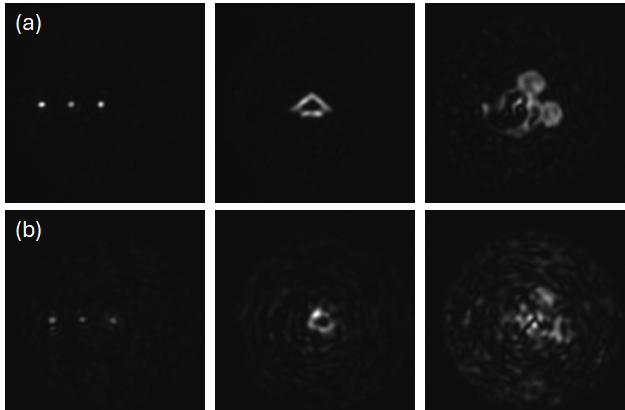}}
\caption{A set of focused spots, a company logo, and a character created in a pedestal-style fiber in the (a) horizontal polarization and (b) vertical polarization created with the scalar transmission matrix. Size of images is 1.1\,mm, which is approximately a 13x magnification of the fiber endface. Fiber calibrated with 35$\times$35 scanned spots.}
\label{fig:PolarizationLogos}
\end{figure*}

As mentioned in Sect.\,\ref{sect:overview}, separately launching and measuring both polarizations effectively doubles the dimensions of the launch coefficient vector $c^{\mathrm{launch}}$ and the distal output E-field vector $E^{\mathrm{out}}$, so that we then have $c^{\mathrm{launch}}\in\mathbb{C}^{2N_{\mathrm{basis}}}$ and $E^{\mathrm{out}}\in\mathbb{C}^{2N_{\mathrm{pix}}}$. While the order of the entries in both vectors is arbitrary (e.g., sorting them such that neighboring entries correspond to different polarizations), it seems advantageous to first put all entries for one polariation, followed by all entries for the other polarizations, because this gives the transmission matrix $T\in\mathbb{C}^{2N_{\mathrm{pix}}\times 2N_{\mathrm{basis}}}$ a 2$\times$2 block matrix structure consisting of four $N_{\mathrm{pix}}\times N_{\mathrm{basis}}$ blocks. The on-diagonal blocks describe the output in a polarization when the same polarization was launched, and the off-diagonal blocks describe the output in one polarization when the other polarization was launched.

We test our polarization control by measuring the full polarization-resolved TM in a pedestal fiber and creating logos rotated by 90 degrees simultaneously in the output horizontal and vertical polarizations as shown in Fig.\,\ref{fig:PolarizationLogos}. The fiber contains a central core that is single-moded at 1064\,nm, but we calibrate the entire surrounding pedestal, which has a diameter of 70\,µm and an NA of 0.13, by choosing the $N_{\mathrm{pix}}$ pixels such that they span the corresponding larger part of the distal camera images. The complete lack of cross-coupling between the rotated logos in Fig.\,\ref{fig:PolarizationLogos} confirms the high fidelity of the polarization-resolved measurement and electric field control. 

For comparison, we also compute the scalar TM of the same fiber and create the same distal patterns in both the horizontal and vertical polarizations. Although slightly dimmer, the spots are re-created in the non-measured polarization with excellent fidelity. Further, the company logo and character are also visible, albeit with degraded quality. That is still remarkable given the accuracy of the TM typically required to create extended distal images, and indicates the polarization properties of this fiber may allow for simplification in its real world use as it may be possible to dispense with the distal polarization optics entirely and still create spots with acceptable SNR for certain imaging applications. We also notice similar behavior from the 200\,µm HCS fiber and the pump-signal combiner operating in both the forward and backwards directions presented in the previous section. We have not observed any such behavior in any of the 50\,µm fibers we have tested, nor, interestingly, in the PM version of the pedestal fiber.\cite{Ahmad2020} 

\section{Far-Field Results}
\label{sec:farFieldResults}

\begin{figure*}
\centerline{\includegraphics[width=\textwidth]{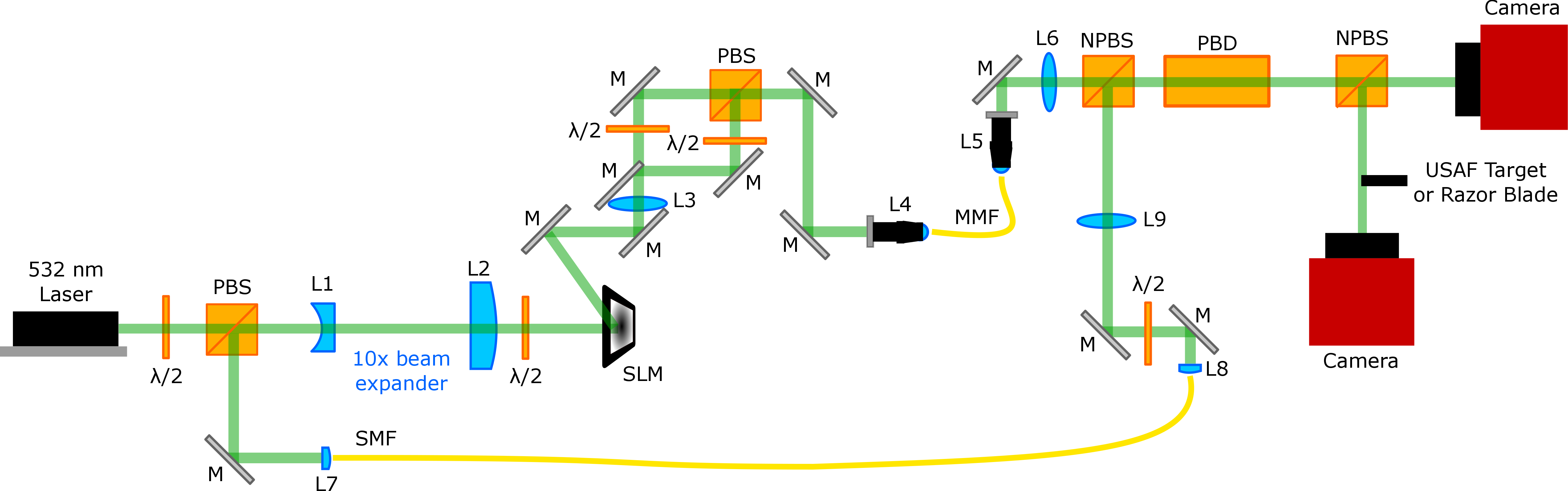}}
\caption{Experimental setup for conducting razor blade contrast and USAF target imaging experiments in the far field. $\lambda$/2: half-wave plate; PBS: polarizing beam splitter; L: lens; SLM: spatial light modulator; M: mirror; MMF: multimode fiber; NPBS: non-polarizing beam splitter; PBD: polarizing beam displacer; SMF: single mode fiber. The lenses used are as follows: L1: -20\,mm achromat; L2: 200\,mm achromat; L3: 400\,mm achromat; L4/L5: 20x microscope objectives; L6: 150\,mm achromat; L7/L8: 11\,mm asphere; and L9: 200\,mm achromat.}
\label{fig:ExperimentFarField}
\end{figure*}

We perform additional experiments with our setup by placing a razor blade or a negative USAF target in the far-field of the distal end of the MMF. The setup for these experiments is shown in Fig.\,\ref{fig:ExperimentFarField}. The main difference from our previous experiments is in the position of the camera used to measure the TM; it has now been moved around 11\,cm past the image plane of the distal facet of the MMF. Although these experiments can also be performed in the image plane of the MMF's distal endface, we achieve clearer images in the far-field due to the improved uniformity of the distal spot focusing in the far field. \cite{Collard2021}

We measure the TM and create a grid of 49x49 distal spots using a variety of multiplexing methods that we then scan over the edge of the razor blade (or USAF target). We collect signal light on a camera placed directly behind the razor blade or target and normalize the power collected to the power in the distal spots. For the razor blade, we calculate the contrast as
\begin{equation}
contrast = \frac{\frac{1}{N_{\mathrm{all}}-N_{\mathrm{blocked}}}\sum\limits_{k\in S_{\mathrm{all}}- S_{\mathrm{blocked}}}(P_{k}-P_{\mathrm{background}})}{\frac{1}{N_{\mathrm{blocked}}}\sum\limits_{k\in S_{\mathrm{blocked}}}(P_{k}-P_{\mathrm{background}})},\label{eq:razorBladeContrast}
\end{equation}
where $N_{\mathrm{all}}$ is the total number of calibrated distal spots (here: $N_{\mathrm{all}}=49^{2}=2401$), $S_{\mathrm{all}}=\{1,\ldots,2401\}$ is the set of all spot indices, $N_{\mathrm{blocked}}$ is the empirically determined number of spots blocked by the razor blade, $S_{\mathrm{blocked}}$ is the set of blocked spot indices and has totally $N_{\mathrm{blocked}}$ integer elements, $P_{k}$ is the power of spot $k$ integrated over the collection camera, and $P_{\mathrm{background}}=N_{\mathrm{pix}}I_{\mathrm{background}}$ is the approximate background power, which is the product of the number $N_{\mathrm{pix}}$ of pixels and the near-maximum grayscale intensity value $I_{\mathrm{background}}$ of a blank camera image. Equation~(\ref{eq:razorBladeContrast}) is thus taking the average collected power of the spots transmitted by the razor blade and dividing by the average of the power collected by the camera for the spots blocked by the razor blade. The SNR of the razor blade image is also calculated as described in the supplementary information of Reference [\onlinecite{Leite2021}].

\begin{figure}
\centerline{\includegraphics[width=3.25in]{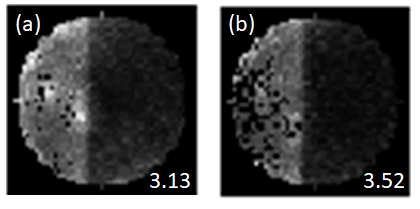}}
\caption{(a) Image of a razor blade with contrast 3.13 in the far field of the distal endface, scanned with 49x49 distal spots created with multiplexing method (a).\cite{Davis1999} (b) Image of a razor blade with contrast 3.52 in the far field of the distal endface, scanned with 49x49 distal spots created with multiplexing method (c).\cite{Arrizon2007}}
\label{fig:Razor}
\end{figure}

Figure~\ref{fig:Razor} shows two examples of images of a razor blade edge taken in the setup shown in Fig.\,\ref{fig:ExperimentFarField}. It should be noted that the contrast and SNR results can depend significantly on the detector used and the processing of the data. We have chosen to use our camera as the power detector due to its greater information capacity than a point detector such as a photodiode or photomultiplier tube. However, the background noise from the camera sensor can have a large effect on the numerical interpretation of the data. We have chosen to subtract a fixed grayscale value $P_{\mathrm{background}}$ from all of the camera images used to calculate powers, with that grayscale value chosen close to the maximum grayscale value seen in a blank image. Otherwise, a small residual background integrated over the large number of camera pixels under consideration can have an out-sized influence on the results. Handling the background subtraction in a different manner changes the numerical contrast and SNR values but tends not to change the overall trends. The usage of the camera as a detector here also results in some non-uniformity of the power in the distal spots even after a normalization step, and results in the dark pixels in the bright portion of the razor blade images due to some distal spots being excluded due to poor data at that point.   

\begin{figure}
\centerline{\includegraphics[width=3.5in]{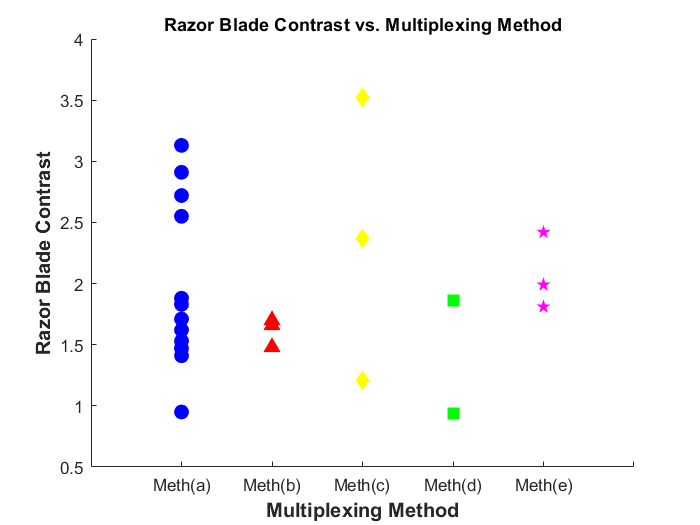}}
\caption{Calculated razor blade contrasts versus tested multiplexing method giving a representative range of the achieved results.}
\label{fig:RazorBladeContrast}
\end{figure}

A representative sample of the razor blade contrasts we achieve is shown in Fig.\,\ref{fig:RazorBladeContrast}. We aim to present the typical result of our experiment as opposed to the best-possible result to indicate the expected quality in a more real-world and day-to-day setting. The SNRs of the razor blade images from these data sets achieve a maximum of over 3.5, which is on par with the results presented in Reference [\onlinecite{Leite2021}]. Although we find the condition number of the TM to be useful in optimizing certain aspects of aligning the experiment, we found that it is not the best predictor of the image quality achieved with distal spots produced from a given TM measurement. We have observed that there are details of the proximal alignment and laser noise that are not captured by the condition number that are more important to the achieved razor blade contrast. For example, re-setting the proximal alignment of the fiber using our alignment strategy discussed in Sect.\,\ref{sect:experimentalDesign} tends to yield higher contrasts than those achieved immediately before re-alignment. Also, measuring the razor blade contrast immediately after the TM measurement produces better results as the validity of the calibration will degrade with elapsing time due to changing environmental conditions within the laboratory. A data set exhibiting particularly high phase drift will also produce lower-contrast imaging results. This chart is intended to show the natural variation in these types of experiments in a laboratory with minimal environmental controls to show the variation that may be expected in real-world applications.

Importantly, most of the tested multiplexing methods can achieve similar razor blade contrasts. It is especially noteworthy that the crudest but most power efficient multiplexing method we tested (method (e)), which combines the launch conditions with the correct phase but all with the same amplitude, yields comparable contrasts to the more precise multiplexing methods, making it a good choice if computational and/or overall power efficiency are a priority in a given application.

\begin{figure*}
\centerline{\includegraphics[width=5in]{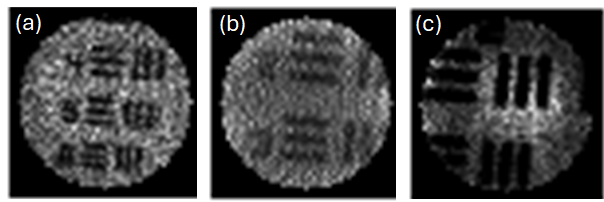}}
\caption{Examples of imaging the negative USAF target through (a) The entire core region of the 200\,µm HCS fiber, (b) the pump-signal combiner in the reverse direction, and (c) the pedestal fiber. The target was imaged with a grid of 45x45 distal spots, with the signal collected by a camera placed after the negative target.}
\label{fig:USAFTargets}
\end{figure*}

We also use focused distal spots created through a variety of our larger core fibers in the far-field to image a negative USAF target and present a few results in Fig.\,\ref{fig:USAFTargets}. The contrast in Fig.\,\ref{fig:USAFTargets}(a) through the 200\,µm HCS fiber is noteworthy given that the TM of this fiber was under-sampled by over a factor of 78, as noted in Sect.\,\ref{sect:alternateFibers}. This indicates that the faster and less computationally intense calibration of these larger core fibers is likely sufficient for many imaging applications.

\section{Discussion and Conclusion}

We have outlined in detail the experimental setup and procedure for measuring a fully polarization-resolved TM and using it to control the distal electric field. We have drawn attention to details that we found important in achieving success in implementing these techniques. As highlighted by the references, there are many variations on the procedure and technology described in this manuscript that can also be used to measure TMs.

We demonstrate the utility of our setup across a variety of specialty fibers, both indicating that the setup works in a variety of fibers without need for re-optimization and that the TM framework is broadly applicable across optical fibers and the applications that they serve. We used fibers with harsh-environment coatings, pedestal fibers, large-core hard-clad silica fibers used in power delivery applications, a rectangular-core fiber, and a pump-signal combiner to illustrate different points about our experiments. We also show that the large core fibers have a similar TM in both of their polarizations, which may allow for simplification of the distal optics in a real application, and we also demonstrate that under-sampling the TM by a factor of 78 still allows for USAF target imaging with our distal spots, providing a route to faster and less computationally expensive calibrations. 

We have focused here on measuring one-way TMs with access to the distal end of the MMF, but the same methods and data processing can be applied to measuring round-trip TMs at the proximal end of the fiber by imaging the light returning to the proximal end of the fiber in place of or in addition to the one-way distal light. Thus, this platform can be used to study and develop fibers for ultrathin medical endoscopes, as well as techniques for enabling proximal-only imaging to expand the use of these devices to real-world settings. 

\section*{Data Availability Statement}

The data that support the findings of this study are available from the corresponding author upon reasonable request.

\section*{REFERENCES}

\nocite{*}
\bibliography{TMTutorial}

\end{document}